\documentclass[11pt]{article}
\usepackage[dvips]{graphicx}

\begin{document}

\title{Minimizing the effect of sinusoidal trends in detrended fluctuation analysis}
\author{ Radhakrishnan Nagarajan \\
Univ. of Arkansas for Medical Sciences \\ Little Rock, Arkansas
72205 \\
Email : nagarajanradhakrish@uams.edu \\ \\
Rajesh G. Kavasseri \\
Dept. of Electrical and Computer Engineering\\
North Dakota State University, Fargo, ND - 58102\\
Email : rajesh.kavasseri@ndsu.nodak.edu}

\date{}

\maketitle

\begin{abstract}
\noindent The detrended fluctuation analysis (DFA) [Peng et al.,
1994] and its extensions (MF-DFA) [Kantelhardt et al., 2002] have
been used extensively to determine possible long-range correlations
in self-affine signals. While the DFA has been claimed to be a
superior technique, recent reports have indicated its susceptibility
to trends in the data. In this report, a smoothing filter is
proposed to minimize the effect of sinusoidal trends and distortion
in the log-log plots obtained by DFA and MF-DFA techniques.
\end{abstract}

\section{Introduction}
There have been several instances where data recorded from a wide
variety of systems exhibit broad-band power law decay
[Bassingthwaite et al, 1995]. These include the class of
self-similar (self-affine) data sets that lack a well-defined
temporal scale. The nature of the decay is reflected by the scaling
exponent  . Such self-similar data sets can be broadly classified
into monofractal and multifractals [Stanely et al., 1999]. While the
former is characterized by a single scaling exponent ($\alpha$ ),
the latter has a spectrum of scaling exponents. Identifying the
scaling exponents has been found to be of practical value in
distinguishing the various states of activities [Ivanov et al.,
1999]. Thus developing suitable techniques to accurately capture the
scaling exponents has been an area of active research. Peng et al.
(1994) introduced the detrended fluctuation analysis (DFA, Appendix
(A)) and demonstrated its superiority over other traditional
techniques such as re-scaled range and Hurst analysis in estimating
the scaling exponent. A scaling exponent $(0 <  \alpha < 0.5)$ is
characteristic of anti-persistent behavior, whereas that of $(0.5 <
\alpha < 1)$ indicates persistent behavior or long-range
correlations in the data. Kantelhardt et al. (2002) extended the DFA
technique (MF-DFA, Appendix (A)) to capture the spectrum of
exponents  in the case of multifractal data. The ease of
implementation and interpretation of results obtained from DFA and
MF-DFA has enjoyed popularity among a wide spectrum of researchers
from diverse disciplines. However, recent studies [Kanterlhardt et
al., 2001; Hu et al., 2001] have pointed out the susceptibility of
the (DFA) to trends in the data. \\

\noindent Kantelhardt et al, (2001) suggested that DFA-$k$ which
incorporates the $k$ 'th order polynomial detrending has been found
to be immune to polynomial trends lesser than $k$. While the
procedure lends itself to be robust to polynomial trends this not
true for other types of trends. Periodic trends are quite common in
experimental data sets, such as diurnal cycles, temperature
fluctuations and seasonal effects. The traditional DFA [Peng et al.,
1994], assumes that the scaling of the fluctuation $F(s$) versus the
time scale $(s$) is in the form of a power law fashion with a single
exponent, i.e. $F(s) \sim s^{\alpha}$ , where the scaling exponent
$(\alpha )$ is estimated by a linear regression of the log-log plot.
There have been instances where the log-log plot exhibit more than
one scaling region. Such a phenomenon has been termed as a crossover
[Peng et al., 1995] and may be used as a pre-cursor to identifying
possible multifractal structure. Hu et al, (2001) pointed out
spurious crossovers in the log-log fluctuation plots which can arise
due to of sinusoidal trends. Thus it is important to develop
techniques to understand whether the crossover is due to the
existence of multiple scaling exponents in the dynamics or a
manifestation of the extrinsic sinusoidal trend. As we shall
demonstrate, polynomial detrending of finite order might not be
useful in removing the sinusoidal trends. This is to be expected as
the Maclaurin's series expansion of a sinusoidal function results in
infinite terms. The present study addresses the above problem and
proposes a combination of smoothing filter in the frequency domain
and $q$'th generalized moment estimation to overcome the above
problem. While the former minimizes the effect of the crossovers
introduced by the sinusoidal trend, the latter provides a
qualitative understanding of the nature of the crossover.

\section{Minimizing the effect of sinusoidal trends}
Sinusoidal trends superimposed on a broad-band power-law noise can
be identified by characteristic peaks in the power spectrum. A
simple averaging technique in the frequency domain can be useful in
eliminating the effect of trends while retaining the power-law
behavior of the noise.  A smoothing filter to accomplish this is
explained below. \\

{\em Smoothing Filter} :

\begin{enumerate}
\item Let the power-law noise be represented by $x$ and the sinusoidal
trend by $t$. Therefore the noise with the trend is given by $y = x
+ t$; with Fourier transform $F(f)$. \item Let the frequency in the
Fourier transform corresponding to the periodic trend occur at
 $f = f_k$.
\item Replace the power at the frequency $f_k$ by a smoothing filter

\begin{eqnarray*}
|F^{\ast}(f_k)| = 0.5(|F(f_{k-1})| + |F(f_{k+1})|)\\
|F^{\ast}(f)| = |F(f)|  \;\;, f \ne f_k
\end{eqnarray*}

\noindent Assign random phases so as to satisfy the conjugacy
constraints. \item  Determine the inverse Fourier transform to
obtain the filtered data $y^f$.

\end{enumerate}

\noindent If the objective is to minimize the power contributed by
the trend, randomizing the phases (Step 4) might not be necessary as
the power spectrum and hence, the auto-correlation is immune to the
phase information in the signal (Weiner-Khintchine theorem) [Proakis
and Manolakis, 1992]. In the following case studies, we shall
utilize the smoothing filter to minimize the effect of sinusoidal
trends on the DFA and MF-DFA estimation procedures. We consider
synthetic and real world data sets.

\subsection*{Case (i) Monofractal long- range correlated noise
corrupted with sinusoidal trends}
 In [Hu et al., 2001], the
susceptibility of the DFA (1) to the monofractal data infected with
sinusoidal trend was discussed. It was concluded that the plot of
$log(F(s))$ vs $log(s)$ exhibited a crossover proportional to the
time period of the sine wave. To determine the effectiveness of the
smoothing filter we use the data published in [Hu et al., 2001]. The
power spectrum of a power-law noise superimposed by a sinusoidal
trend with amplitude $A = 2$ and period $T = 2^7$ is shown in Fig
1a, Appendix (B). The DFA of the power-law noise for the various
order polynomial detrending is shown in Fig 1b. As expected the
scaling exponent from the log-log plot is   $\alpha = 0.9$,
conforming to earlier reports [Hu et al., 2001]. However, the
log-log plot of the power-law noise with sinusoidal trend exhibits a
characteristic crossover, Fig 1c. It is important to note that the
crossover is due to the trend and not the dynamics. Despite higher
order polynomial detrending $(d = 1, 2, 3, 4, 5$, and $6)$, the
spurious crossover induced by the sinusoidal persists, see Fig 1c.
The distortion introduced by the sinusoidal trend imparts a
nonlinear structure to the fluctuation function and prevents the
application of linear regression to estimate the scaling exponent.
However, the fluctuation plots obtained after applying the smoothing
filter resembles straight lines with ($\alpha  \sim 0.9)$, Fig 1d,
similar to that of the trend free power law noise, Fig 1b. Thus the
spurious nonlinear structure introduced by the sinusoidal trend is
minimized on applying the smoothing filter prior to the DFA
estimation procedure.

\begin{figure}[htbp]
\centering
\includegraphics[height=4in, width=4in]{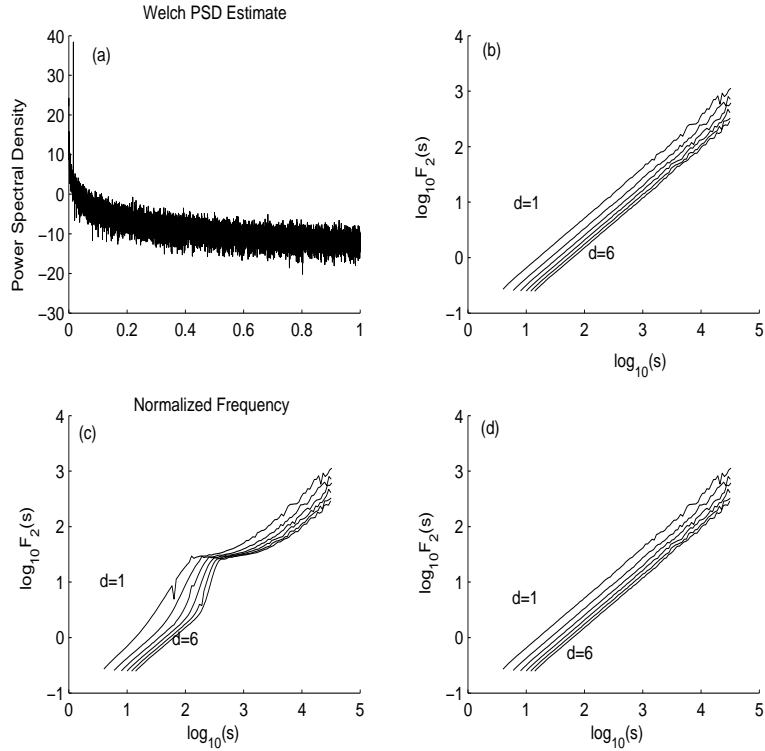}
\caption{Power spectrum of monofractal data $(\alpha  = 0.9)$
corrupted with sinusoidal trend is shown in (a). The log-log plot of
the fluctuations obtained by applying the second-order DFA $(q = 2)$
for the various polynomial detrendings $(d = 1, 2, 3, 4, 5 \;$and$
\;6)$ for the monofractal data, monofractal corrupted with
sinusoidal trend, and that reconstructed using the smoothing filter
is shown in (b, c and d) respectively.} \label{}
\end{figure}

\subsection*{Case (ii) Monofractal long-range correlated noise
corrupted with multiple sinusoidal trends} To determine the effect
of multiple sinusoidal trends, the power-law noise $( \alpha = 0.9)$
was corrupted with sinusoidal trends with parameters $(A_1 = 6, A_2
= 3, A_3 = 2, T_1 = 2^6, T_2 = 2^4; T_3 = 2^2,$ Appendix (B). The
frequencies $(f_i =1/T_i)$ of the trends are chosen such that $f_2$
and $f_3$ represent the second and the third harmonic of the
fundamental $f_1$. Harmonic trends are commonly observed in
experimental data hence its discussion in the present study. The
power spectrum of the power law noise with multiple sinusoidal
trends is shown in Fig 2a. The corresponding log-log plot of the
fluctuation is shown in Fig 2b and resembles that obtained in the
case (i). Similar to case (i), the smoothing filter is useful in
minimizing the effect of the trends introduced by the sinusoidal
trends and thereby facilitates a reliable extraction of the scaling
exponent $(\alpha \sim 0.9)$. The traditional DFA captures only the
second moment $(q = 2$, i.e. $F_2(s))$ similar to power spectral
techniques. To examine the effect of the generalized moments $(q)$,
the fluctuations were computed with polynomial order (d = 4) and
varying $q = (-10, -8, -6, -4, -2, 2, 4, 6, 8, 10)$. The log-log
plot of the fluctuations $F_q(s$) with respect to the time scale $s$
with varying $q$ on the power law noise corrupted with harmonic
trends is shown in Fig 2d. While the log-log plot fails to exhibit a
linear trend, the nature of the fluctuations does not change
appreciably with varying $q$. Thus, in the absence of the smoothing
filter, determining the qualitative behavior of the fluctuation for
various values of $q$ can provide insight into the nature of the
crossover in the given data. Such an analysis can be helpful in the
case of more complex dynamics such as multifractal noise as
discussed below.

\begin{figure}[htbp]
\centering
\includegraphics[height=3.5in, width=4in]{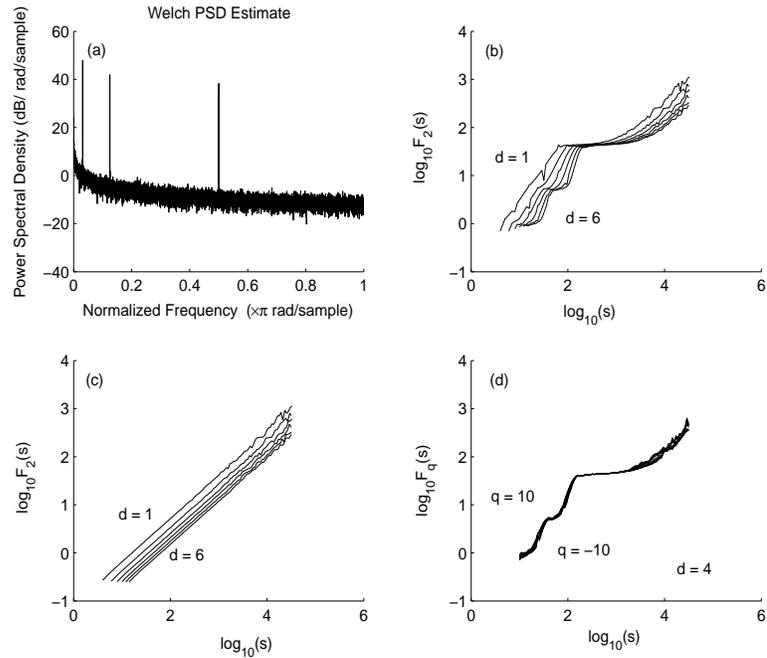}
\caption{Power spectrum of monofractal data $( \alpha = 0.9)$
corrupted with sinusoidal trends is shown in (a). The log-log plot
of the fluctuations obtained by applying the second-order DFA $(q =
2$) for the various polynomial detrendings $(d = 1, 2, 3, 4, 5
\;$and$\; 6)$ for the monofractal corrupted with sinusoidal trend
and that reconstructed by the smoothing filter is shown in (b and c)
respectively. The fluctuation $F_q(s)$ obtained for $(q = -10, -8,
-6, -4, -2, 2, 4, 6, 8, 10)$ with $(d = 4)$ is shown in (d).}
\label{}
\end{figure}

\clearpage

\subsection*{Case (iii) Multifractal noise corrupted with multiple
sinusoidal trends} The multifractal data considered is that of
internet traffic [Levy Vehel and Reidi, 1996]. The data was
corrupted with sinusoidal trends with amplitude and the time period
$(A_1 = 6000, A_2 = 3000, T_1 = 2^6, T_2 = 2^4$, Appendix (C). As in
case (ii), the frequencies are harmonically related to one another.
The MF-DFA of the multifractal data estimated with polynomial order
$(d = 4)$ and varying $q = (-10, -8, -6, -4, -2, 2, 4, 6, 8, 10)$ is
shown in Fig. 3b. The log-log plots exhibit a marked change in the
slope characteristic of multifractal data. The log-log plot of the
multifractal data after applying the smoothing filter is shown in
Fig. 3d and resembles that of Fig 3b. Fig. 3c shows the log-log
fluctuation of the multifractal data with the sinusoidal trend. It
can be observed unlike case (ii), Fig. 2d, the slopes show a
dramatic change with varying $q$. Thus even in the absence of a
smoothing filter the analysis of the log-log with varying $q$ can
provide insight into the nature of the dynamics. However, the
estimation of the slopes for multifractal data is a non-trivial
issue and not discussed in the present study.

\begin{figure}[htbp]
\centering
\includegraphics[height=4in, width=4in]{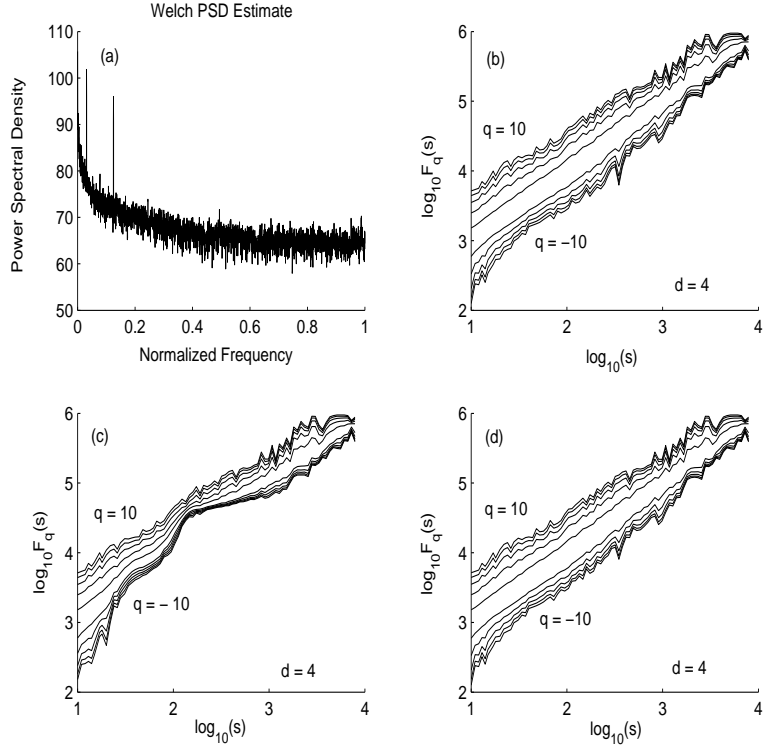}
\caption{Power spectrum of multifractal data corrupted with
sinusoidal trends is shown in (a). The fluctuation $F_q(s)$ obtained
for $(q = -10, -8, -6, -4, -2, 2, 4, 6, 8, 10)$ with $(d = 4)$ for
the multifractal data, multifractal data corrupted with sinusoidal
trends, and that obtained by a smoothing filter is shown in (b, c,
and d) respectively.} \label{}
\end{figure}

\section{Conclusions} The retention of sinusoidal trends despite higher
order polynomial detrending can induce spurious crossovers and
prevent reliable extraction of scaling exponents in DFA/MF-DFA
procedures. In the present report, a simple smoothing filter is
proposed to minimize the effect of sinusoidal trends on the
DFA/MF-DFA procedures.  The effectiveness of the smoothing filter on
monofractal and multifractal data corrupted with sinusoidal trends
were considered. While the smoothing filter minimizes the trends it
is nevertheless an approximation. A qualitative approach to
determine whether the observed distortion in the log-log plots is an
outcome of the intrinsic dynamics or trend is also discussed. It is
important to note that only sinusoidal trends whose power is much
smaller than that of the power-law noise were discussed in the
present study.

\section*{References}

\noindent Bassingthwaighte., J.B. Liebovitch., L.S. West., B.J.
[1995] {\em Fractal Physiology}, Oxford Univ. Press. \\

\noindent Hu., K. Ivanov., P.C.h. Chen., Z. Carpena., P. and Stanley
H.E. [2001] ``Effect of trends on detrended fluctuation analysis,"
{\em Phys. Rev. E.}  {\bf 64}, 011114 (19 pages).\\

\noindent Ivanov., P. Ch. Amaral., L. A. N. Goldberger., A. L.
Havlin., S. Rosenblum., M. G. Struzik., Z. R. and Stanley., H. E.
[1999] ``Multifractality in human heartbeat dynamics," {\em Nature}
{\bf 399},
461-465. \\

\noindent Kantelhardt., J.W. Koscielny-Bunde., E. Rego., H.A.H.
Havlin., S. Bunde., A. [2001] ``Detecting Long-range Correlations
with Detrended Fluctuation Analysis," {\em Physica A}, {\bf 295}, 441-454. \\

\noindent Kantelhardt., J.W. Zschiegner., S.A. Koscielny-Bunde., E.
Havlin., S. Bunde., A. and Stanley., H.E. [2002] ``Multifractal
detrended fluctuation analysis of nonstationary time series,"
{\em Physica A} {\bf 316}, 87. \\

\noindent Levy Vehel., J. and Reidi., R. H. [1996] Fractional
Brownian motion and data traffic modeling: The other end of the
spectrum, {\em Fractals in Engineering}, (Eds. Levy Vehel, J.
Lutton, E.
and Tricot, C.) Springer-Verlag.\\

\noindent Peng., C-K. Buldyrev., S.V. Havlin., S. Simons., M.
Stanley., H.E. Goldberger., A.L. [1994] ``On the mosaic organization
of DNA sequences," {\em Phys. Rev. E.} {\bf 49}, 1685-1689. \\

\noindent Peng., C-K. Havlin., S. Stanley., H.E. Goldberger., A.L.
[1995] ``Quantification of scaling exponents and crossover phenomena
in nonstationary heartbeat time series," {\em Chaos} {\bf 5}, 82-87.\\

\noindent Proakis., G. and Manolakis., D. [1992] ``Digital Signal
Processing, Principles, Algorithms, and Applications", 2nd Ed.
Macmillan.\\

\noindent Stanely., H.E. Amaral., L.A.N. Goldberger., A.L. Havlin.,
S. Ivanov., P.Ch. Peng., C-K. [1999] ``Statistical physics and
physiology: Monofractal and multifractal approaches", {\em Physica
A} {\bf 270}, 309-324.

\newpage

\section*{Appendix A : Multifractal Detrended Fluctuation Analysis (MFDFA)}

A brief description of the Multifractal DFA (MF-DFA) [Kantelhardt et
al., 2002] is included below for completeness of the report. The
scaling exponent in the case of monofractal data can be obtained by
fixing $(q = 2)$.

\begin{enumerate}
\item The integrated series of the given data $\{x_k\}$ is
generated as
\begin{eqnarray*}
y(k) = \sum_{i=1}^{i=k} [x(i) - \bar{x}] \;\;\; k = 1, \dots N
\end{eqnarray*}
where $\bar{x}$ represents the average of $\{x_k\}, k = 1 \dots N$

\item The data is divided in to $n_s$ non-overlapping
boxes of equal length$s$ where $n_s = int(N/s)$. The local
polynomial trend $y_v(i)$ is calculated in each of the bins $v = 1
\dots n_s$ by a polynomial regression and the variance is determined
from

\begin{eqnarray*} F^2(v,s) = \{ \frac{1}{s} \sum_{i=1}^{i=s}
\{y [N-(v-n_s)s + i] - y_v(i)\}^2
\end{eqnarray*}

The order $m$ of the polynomial trend is chosen such that $s \geq m
+ 2$. Polynomial detrending of order m is capable of eliminating
trends up to order $m-1$. This procedure is repeated from the other
end of the data in order to accommodate all the samples. Thus the
effective length of the data is $2 n_s$.

\item The $qth$ order fluctuation function is calculated from
averaging over all segments.

\begin{eqnarray*}
F_q(s) = \left \{ \frac{1}{2 n_s} \sum_{i=1}^{i= 2
n_s}[F^2(v,s)]^{q/2}  \right \} ^{1/q}
\end{eqnarray*}
In general, the index $q$ can take any real value except zero,
[Kantelhardt et.al., 2002]. The scaling behavior is determined by
analyzing the log-log plots $F_q(s)$ versus $s$ for {\em each} $q$.
If the original series $\{x_k \}$ is power-law correlated, the
fluctuation function will vary as
\begin{eqnarray*}
F_q(s) \sim s^{h(q)} \label{hq} \end{eqnarray*}
\end{enumerate}
A change in the shape of the log-log plots with varying $q$ is
indicative of multifractality (Fig. 3).

\section*{Appendix B :Monofractal noise corrupted with sinusoidal trend
$(y)$}

The monofractal data  with scaling exponent $(  \alpha = 0.9)$ was
generated using the algorithm of Makse et al., 1996. This data was
recently used to study   the effect of sinusoidal trends on the
detrended fluctuation analysis. The data is publicly  available at
\newline \noindent http://www.physionet.org/physiobank/database/synthetic/tns/.

\noindent This data was corrupted with sinusoidal trend given by
$t_i(n) =
A_i\sin(2\pi n/T_i),\; n = 1 \dots N$. to obtain $y$. \\

{\bf Data 1 : Synthetic }
\begin{eqnarray*}
y(n) = s(n) + A_1\sin(2\pi n/T_1),\; n = 1 \dots N, \;A_1 = 2, T_1 =
2^7
\end{eqnarray*} shown in Fig.1. \\

{\bf Data 2 : Synthetic}
\begin{eqnarray*}
y(n) = s(n) + A_1\sin(2\pi n/T_1) + A_2\sin(2\pi n/T_2) +
A_3\sin(2\pi n/T_3)
\end{eqnarray*}

with $A_1 = 6,\; A_2 = 3,\; A_3 = 2,\; T_1 = 2^6,\; T_2 = 2^4,\; T_3
= 2^2$, shown in Fig. 2. It should be noted that $T_1 = 16 T_3$ and
$T_2 = 4 T_3$. \\

\section*{Appendix C :Multifractal noise corrupted with sinusoidal trend
$(y)$}

{\bf Data 3 : Real world data} The multifractal data (s) considered
is that of internet  log traffic [Levy et al., 1996].
\begin{eqnarray*}
y(n) = s(n) + A_1\sin(2\pi n/T_1) + A_2\sin(2\pi n/T_2), n = 1 \dots
N, N = 2^{15}.
\end{eqnarray*}
with $A_1 = 6000,\; A_2 = 3000,\; A_3 = 2,\; T_1 = 2^6,\; T_2 =
2^4$.\\

\noindent It should be noted that $T_1 = 4 T_2$ as shown in Fig.3.

\end{document}